\documentclass[11pt]{article}
\usepackage{blois,epsfig}

\def\be{\begin{equation}}
\def\ee{\end{equation}}
\def\bea{\begin{eqnarray}}
\def\eea{\end{eqnarray}}

\begin{document}
\vspace*{4cm}
\title{PRODUCTION OF CHARM AND CHARMONIUM\\
WITH THE ATLAS DETECTOR AT 7 TEV}

\author{ DARREN D.\ PRICE \\
on behalf of the ATLAS Collaboration}

\address{Department of Physics, Indiana University, Bloomington, IN 47405, USA}
\maketitle\abstracts{
We report on the observation of the charm mesons $D^{\star\pm}$, $D^\pm$ and $D^\pm_s$ with 1.4~nb$^{-1}$ of data, and of the $J/\psi\to\mu^+\mu^-$ resonance with
78~nb$^{-1}$ of data from the ATLAS detector in 7~TeV proton-proton collisions at the LHC. The resultant signals
support the high performance of the ATLAS detector as predicted from simulation and prospects for future measurements in the charm and charmonium sector.
}
The observation of charm mesons and $J/\psi$ at ATLAS\,\cite{ATLAS} in early data is useful as both a physics and performance tool. For performance purposes, reconstruction of these states
allows for tests of magnetic field and material maps, detector alignment and Monte Carlo modelling of the detector. 
Reconstruction of $D$ mesons relies on accurate Inner Detector (low-$p_T$) tracking and vertexing, while $J/\psi\to\mu^+\mu^-$ observation additionally relies on the muon
spectrometer system for identification and measurement. Both systems may be adversely affected by detector misalignment and mismodelling effects. 
The ability of ATLAS to reconstruct these states (due in part to their large cross-sections and distinctive decay signatures) and ultimately measure their 
production rates and properties will be a test of theoretical predictions at LHC energies (which currently have large uncertainties and many open questions), 
probes of proton structure functions, and are important steps toward distinguishing and measuring heavy flavour and prompt charmonium contributions to future QCD and B-Physics studies.

\section{Reconstruction of charm mesons}
Data corresponding to a total integrated luminosity of $1.4$~nb$^{-1}$ were selected using a minimum-bias trigger
(with over 99.5\% efficiency for events
with at least two tracks in the beam spot) and after requiring that a reconstructed primary vertex with identified in the collisions. Charm meson candidates
were constructed by combining tracks identified by the ATLAS Inner Detector (ID). To ensure high reconstruction efficiency and good momentum resolution, tracks were required 
to have at least one pixel hit, four hits in the silicon tracker, and be within the fiducial acceptance of the detector. The general selection strategy\,\cite{DmesonCONF} was to utilise the
kinematics of charm production/decay (via cuts on the transverse momenta of the $D$ meson and associated particles in the decay), the hard nature of charm fragmentation
(making cuts on $p_T(D)/\sum{E_T}$, where $\sum{E_T}$ is the sum of the transverse energy of the tracks from the decay), the relatively long $D$-meson lifetimes 
(cutting on the transverse decay length of the $D$) and improving background rejection through the application of cuts on angular properties of the decay.
The signals are measured in the kinematic acceptance range $p_T(D)>3.5$~GeV and $|\eta(D)|<2.1$. 

$D^{\star\pm}$ mesons were reconstructed through the decay channel $D^{\star\pm}\to D^{0}\pi^\pm_s\to (K^\mp\pi^\pm)\pi^\pm_s$. The reconstruction technique 
makes use of the soft pion $\pi^\pm_s$ decay momentum being constrained by the small mass difference between the $K\pi$ and $K\pi\pi_s$ systems. Pairs of oppositely-signed 
tracks, each with $p_T>1.0$~GeV were combined to form $D^0$ candidates, where kaon and pion masses were assumed for each track in turn in making the combinations.
Pairs were kept where the common vertex fit $\chi^2$ value was below 5. In addition, the $D^0$ candidate itself was required to
satisfy conditions on the transverse and longitudinal impact parameters with respect to the primary
vertex, and to have a positive transverse decay length $L_{xy}(D^0)>0$. Combining with an additional track with $p_T>0.25$~GeV (of opposite charge to the kaon) and
assigned a pion mass hypothesis, a $D^{\star\pm}$ is formed. 
\begin{figure}[htbp]
  \begin{center}
    \includegraphics[width=0.4\textwidth]{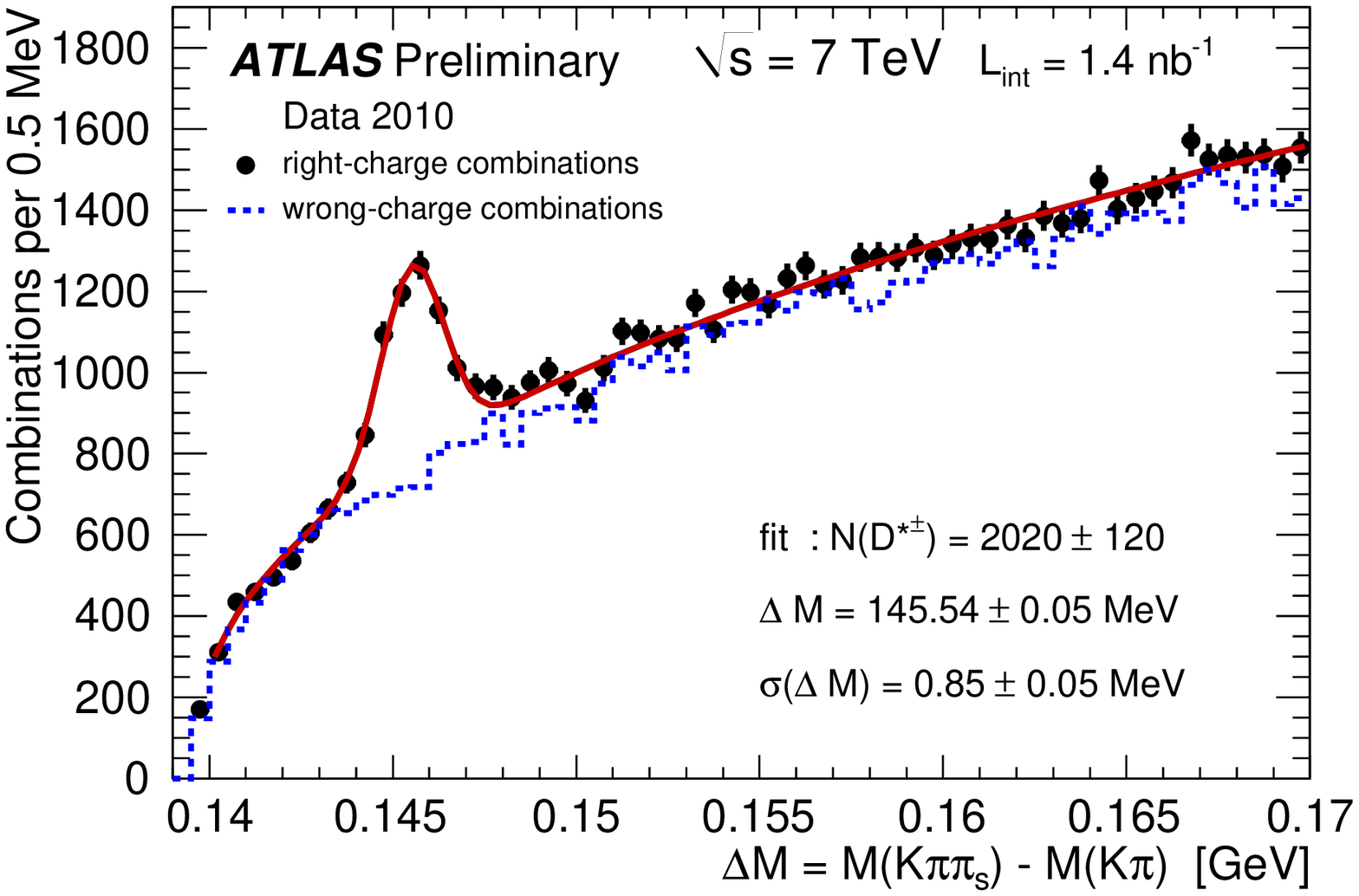}
    \includegraphics[width=0.4\textwidth]{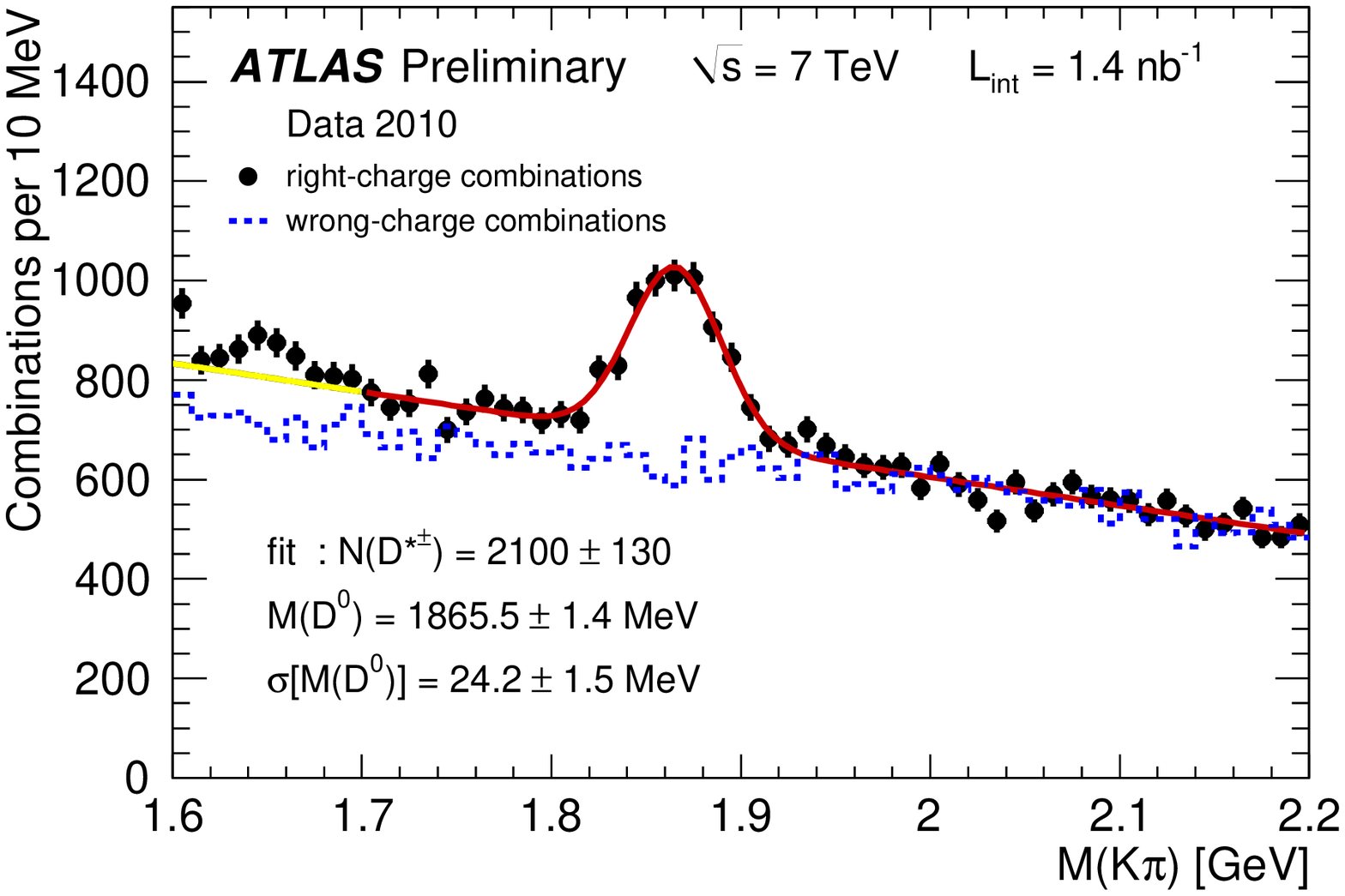}
    \caption{The mass difference $\Delta M = M(K\pi\pi_s) - M(K\pi)$ (left) with $1.83<M(K\pi)<1.90$~GeV mass constraint, showing $D^{\star\pm}$ candidates (points), and the $K\pi$
      invariant mass (right) with a $144<\Delta M<147$~MeV requirement, selecting $D^0$ candidates (points). Dashed histograms show wrong-charge combinations.
      \label{fig:Dstar}}
  \end{center}
\end{figure}
The resultant signals for the $D^{\star\pm}$ and $D^0$ can be seen in Figure~\ref{fig:Dstar} along 
with the wrong-charge combinations, where a constraint on the mass difference $\Delta M = M(K\pi\pi_s) - M(K\pi)$ and $M(K\pi)$ has been applied.
The $D^{\star\pm}$ yields from fits to the $\Delta M$ and $M(K\pi)$ distributions are $2020\pm 120$ and $2100\pm 130$ respectively. The fitted mass and mass differences 
of the $D^{\star\pm}$ and $D^0$ are in agreement with world averages, and the fitted resolutions in good agreement with expectation from simulation.

\begin{figure}[htbp]
  \begin{center}
    \includegraphics[width=0.4\textwidth]{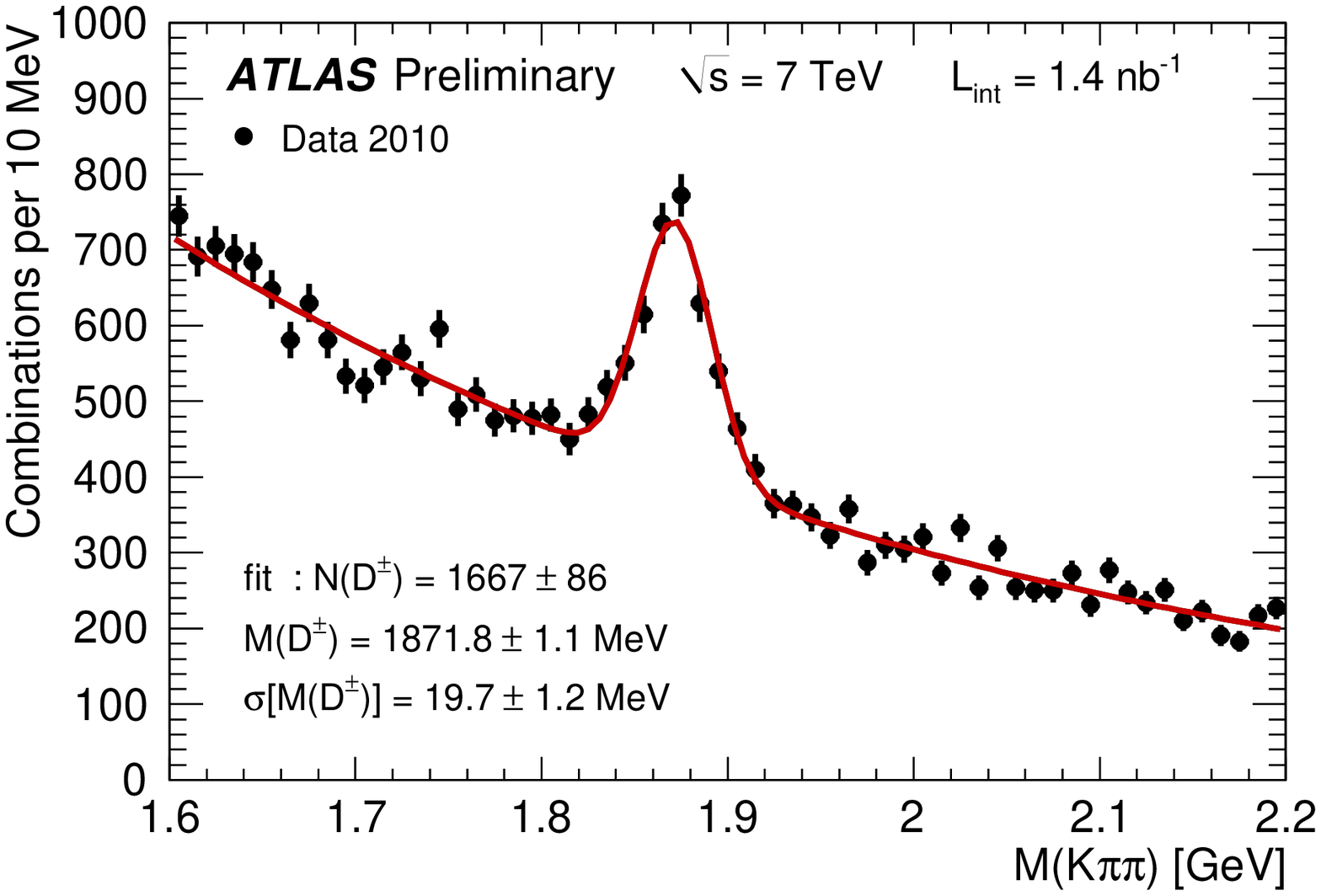}
    \includegraphics[width=0.4\textwidth]{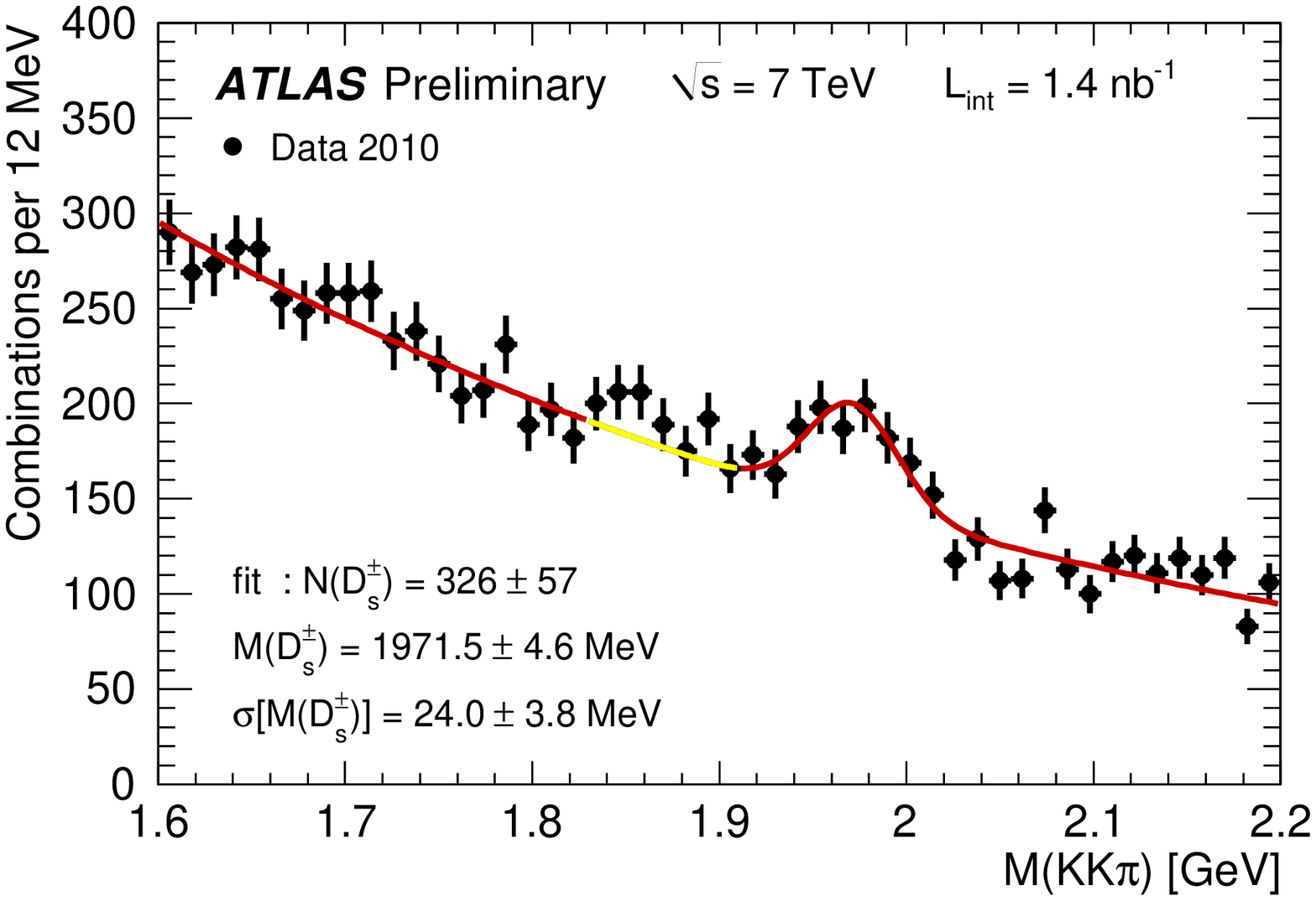}
    \caption{The $M(K\pi\pi)$ distribution for $D^\pm$ candidates (left) and $M(KK\pi)$ distribution for $D^\pm_s$ candidates (right).
      \label{fig:Dmesons}}
  \end{center}
\end{figure}
Reconstruction of $D^\pm$ mesons proceeds using a similar technique, through the decay of $D^\pm\to K^\mp\pi^\pm\pi^\pm$. Two same-charge tracks with $p_T>0.8,1.0$~GeV were
combined with a track of opposite sign and $p_T>1.0$~GeV to form the $D^\pm$ candidate. The three fitted tracks were required to pass fit quality cuts and primary vertex
pointing constraints, and the measured transverse decay length of the candidate was required to be greater than 1.3~mm. The analysis made use of the angular distribution 
between the kaon in the $K\pi\pi$ rest frame and $K\pi\pi$ direction of flight in the lab frame to provide discrimination to combinatorial backgrounds. To suppress
reflections from $D^{\star\pm}$ decay, combinations with $M(K\pi\pi) - M(K\pi)<0.15$~GeV were excluded. $D^{\pm}_s\to\phi(\to K^+K^-)\pi^\pm$ reflections were suppressed
by rejecting $D^\pm$ candidates where, under a kaon mass hypothesis for the tracks, an oppositely-charged track-pair in the candidate would be within 8~MeV of the $\phi$ mass.
Figure~\ref{fig:Dmesons} (left) shows the $D^\pm$ signal after cuts, with a Gaussian fit for the signal and exponential function for the non-resonant background. 
$D^\pm_s$ mesons were reconstructed through the decay $D^{\pm}_s\to\phi(\to K^+K^-)\pi^\pm$, again combining two same-charge tracks ($p_T>0.7$~GeV) and assigned a kaon mass
and combined with a third, oppositely-signed track, of $p_T>0.8$~GeV to form the $D^\pm_s$ candidate. The candidate is kept if the $M(KK)$ invariant mass was within 6~MeV
of the $\phi$ mass. Vertex fit, pointing requirements and transverse decay length cuts are again applied to the candidate, and additionally the angles between the
$\pi$ and the $D^\pm_s$, and the $\phi$ and the $D^\pm_s$ are used to suppress combinatorial backgrounds. The $D^\pm_s$ signal can be seen in Figure~\ref{fig:Dmesons} (right).
The fitted masses for both the $D^\pm$ and $D^\pm_s$ are in agreement with the world average, and width of the signals are in line with expectation from simulation.

\section{Reconstruction of the \boldmath{$J/\psi$} resonance}

Events were selected by either of a minimum bias trigger requiring the presence of a muon at the high-level trigger stage, or by a single-muon hardware trigger
(see references \cite{jpsiCONF1} and \cite{jpsiCONF2} for more details), the data-sample corresponding to an integrated luminosity of 78~nb$^{-1}$. 
In addition, the events were required to contain at least one primary vertex built from three tracks, with each track having at least one hit in the pixel detector and six in the silicon tracker. 
Pairs of oppositely-signed muon candidates are constructed, both of which must have associated ID tracks passing the above selection. No additional cuts are imposed on the muon,
but the muon must have had a total momentum greater than 3~GeV and $|\eta|<2.7$ in order to have been identified by the muon spectrometer. 
The ID tracks of the two muons are refitted to a common vertex, and refitted track parameters and covariance matrices are used to calculate the invariant mass 
and per-candidate mass error.

\begin{figure}[htbp]
  \begin{center}
    \includegraphics[width=0.32\textwidth]{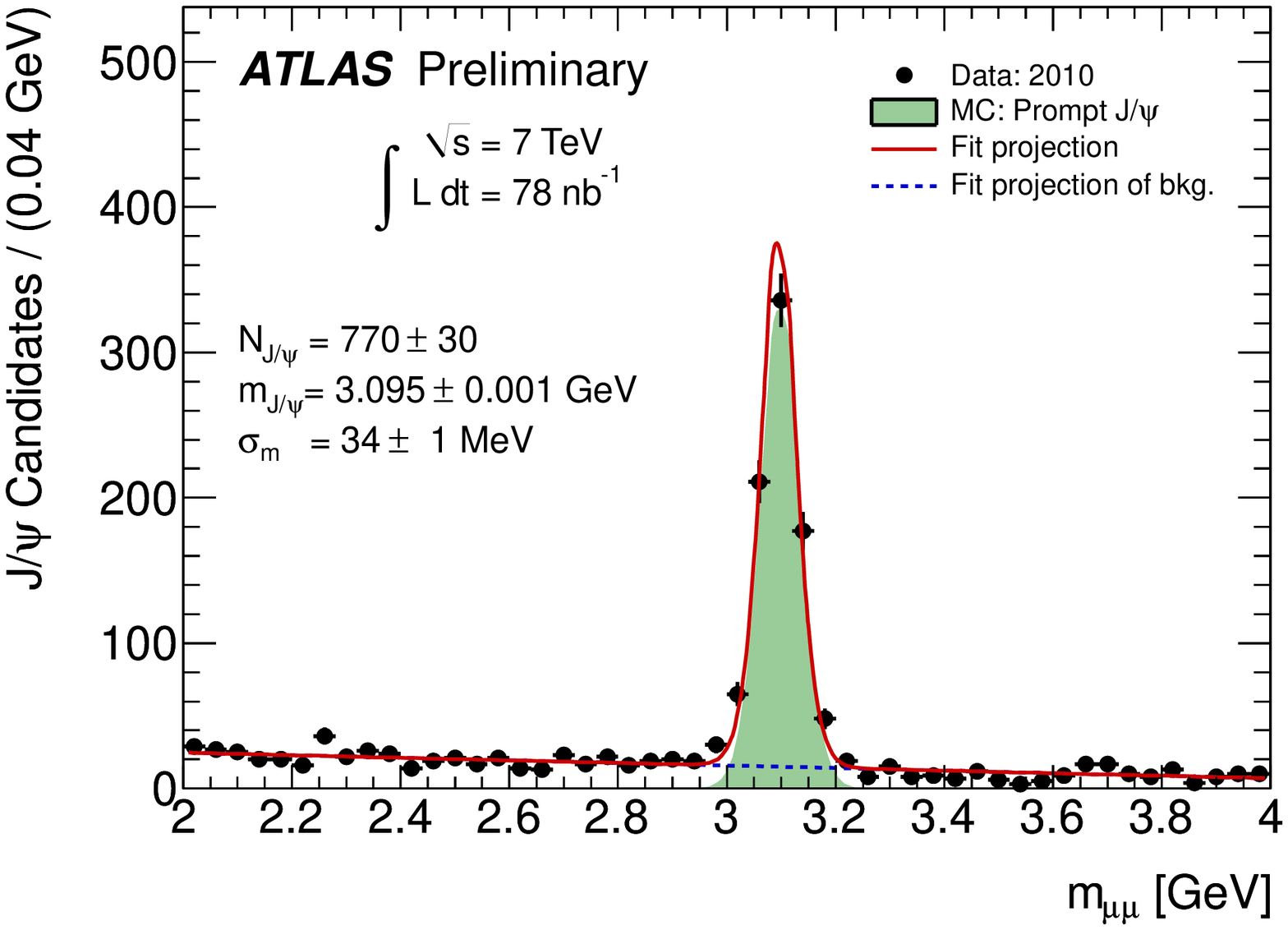}
    \includegraphics[width=0.32\textwidth]{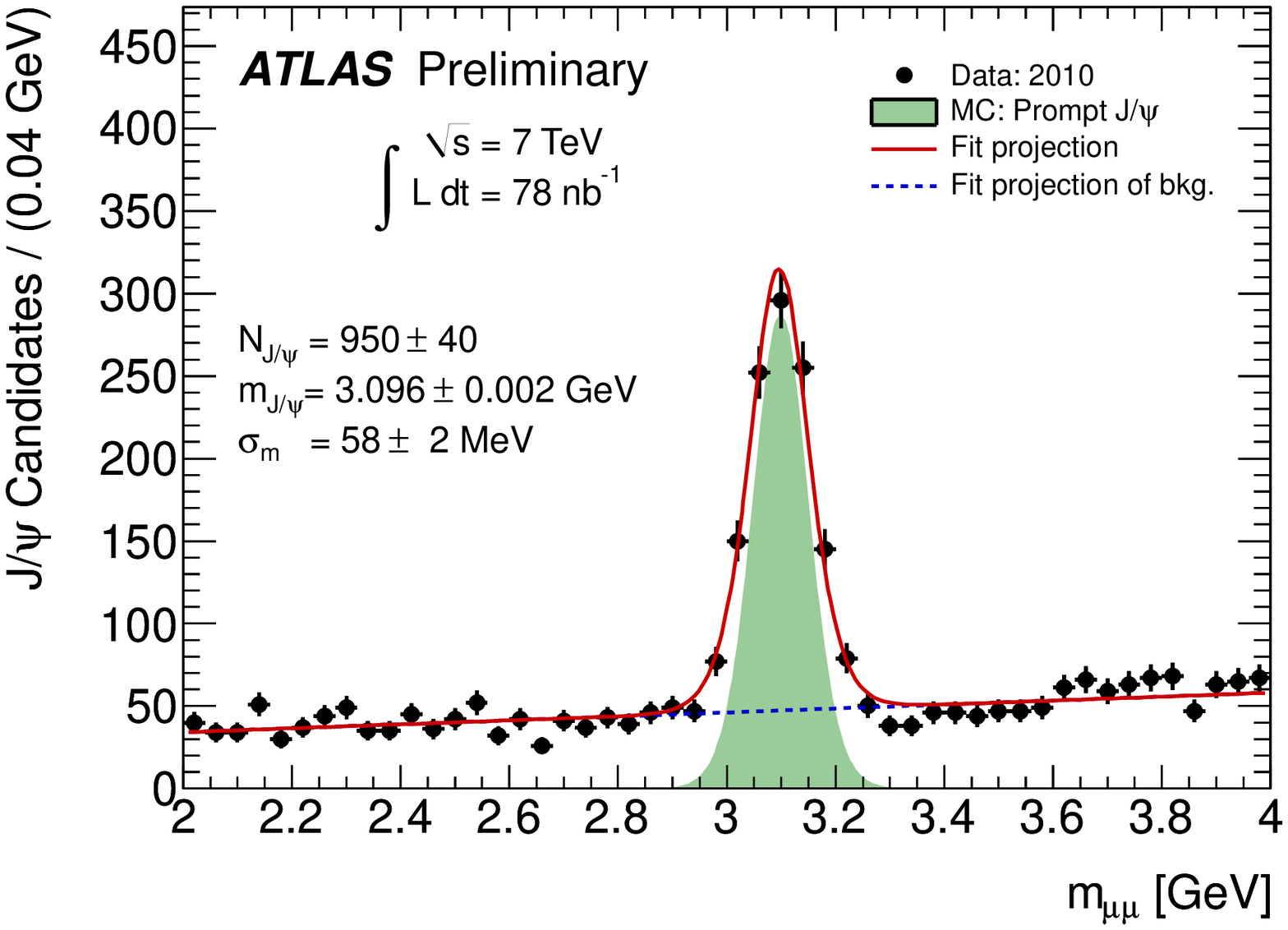}
    \includegraphics[width=0.32\textwidth]{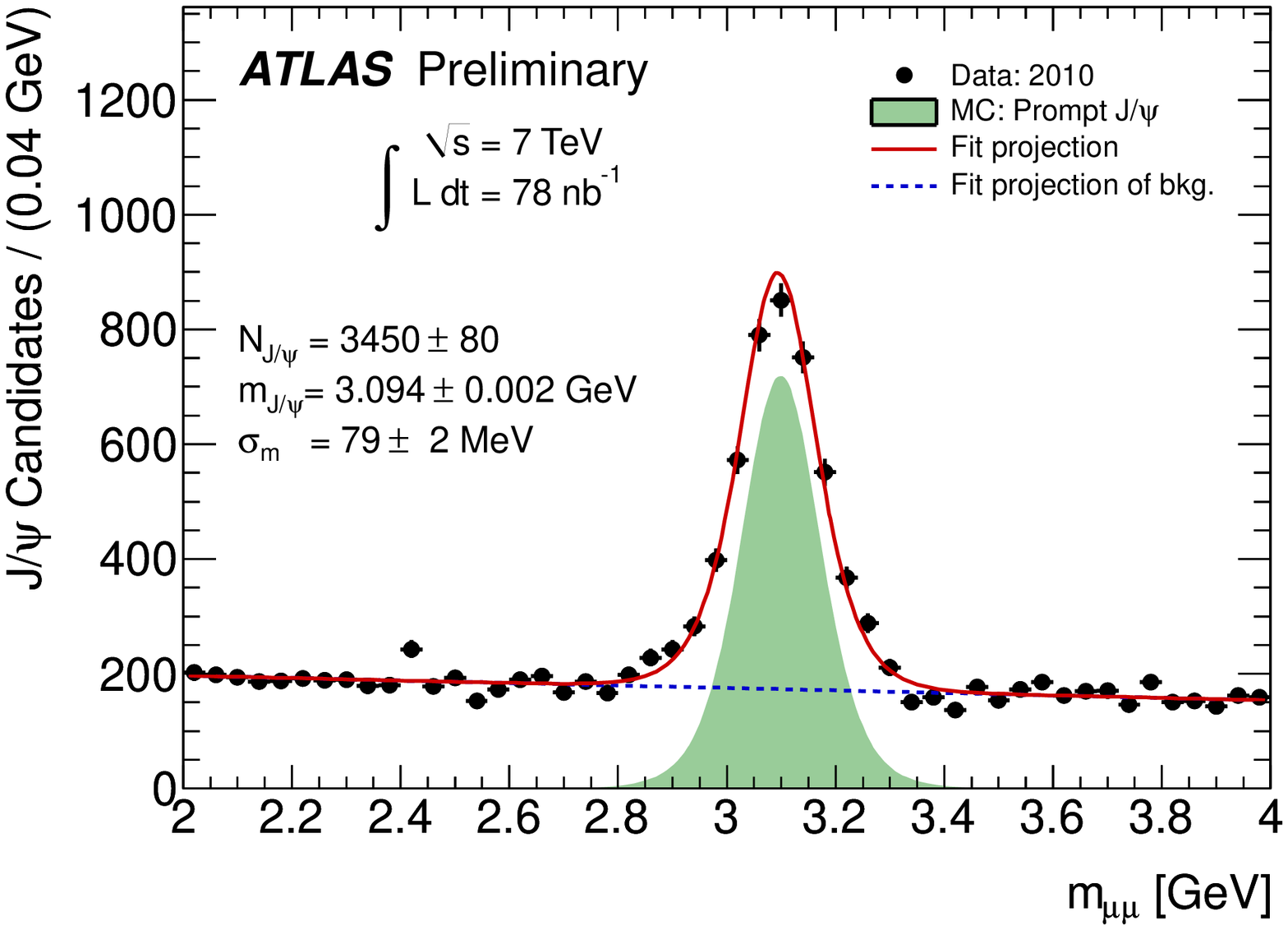}
    \caption{Invariant mass distributions of $J/\psi\to\mu^+\mu^-$ candidates where both muons are found in the barrel region ($|\eta_\mu|<1.05$) [left], one is found in the barrel, 
      the other in the endcap region ($|\eta_\mu|>1.05$) [centre], or where both are found in the endcap [right]. The solid line represents the results of the unbinned maximum likelihood
      fit to the candidates, the dashed line displays the projection of the background-only projection of the same fit. 
      \label{fig:jpsimass}}
  \end{center}
\end{figure}
\begin{figure}[htbp]
  \begin{center}
    \includegraphics[width=0.35\textwidth]{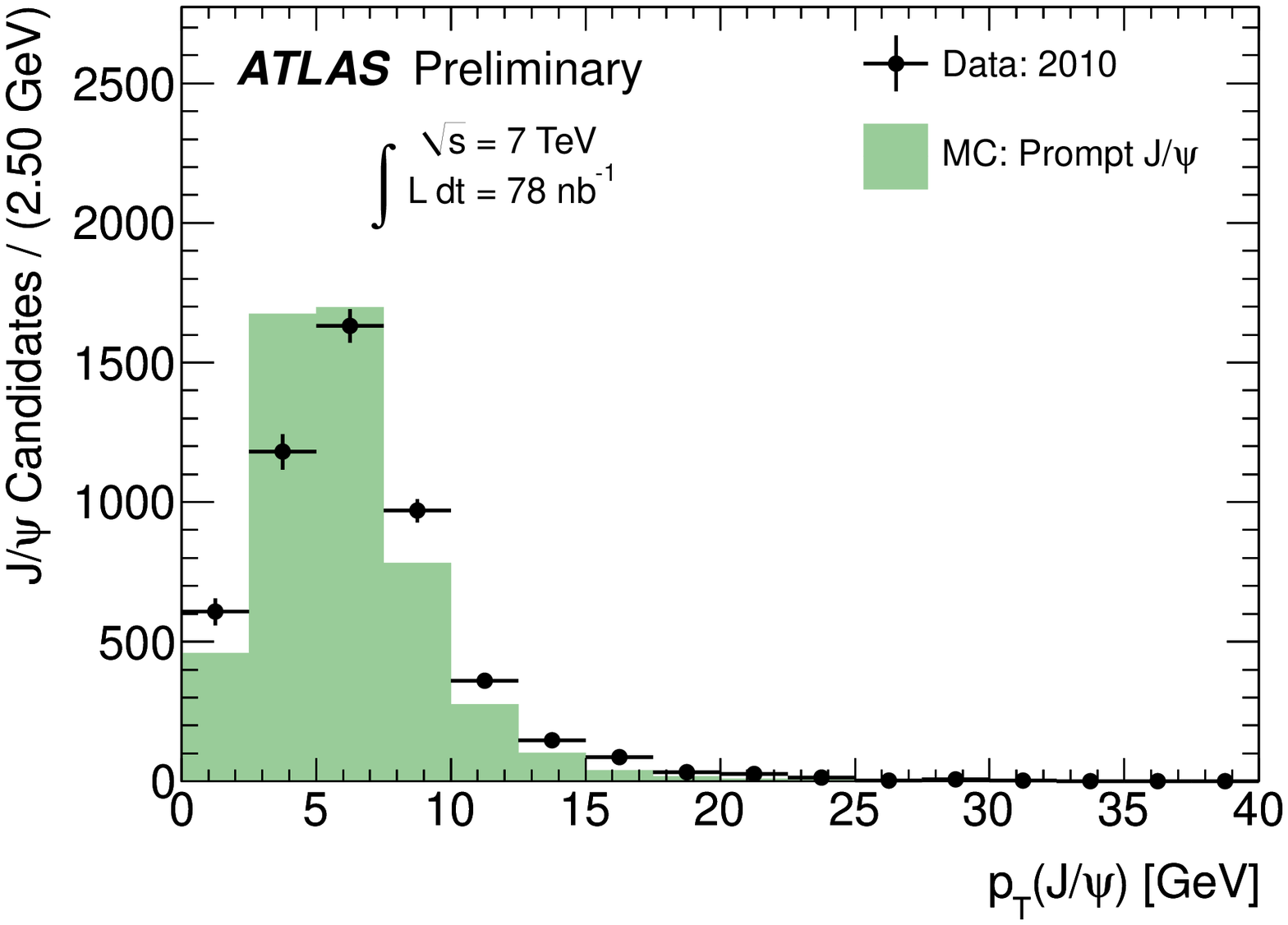}
    \includegraphics[width=0.35\textwidth]{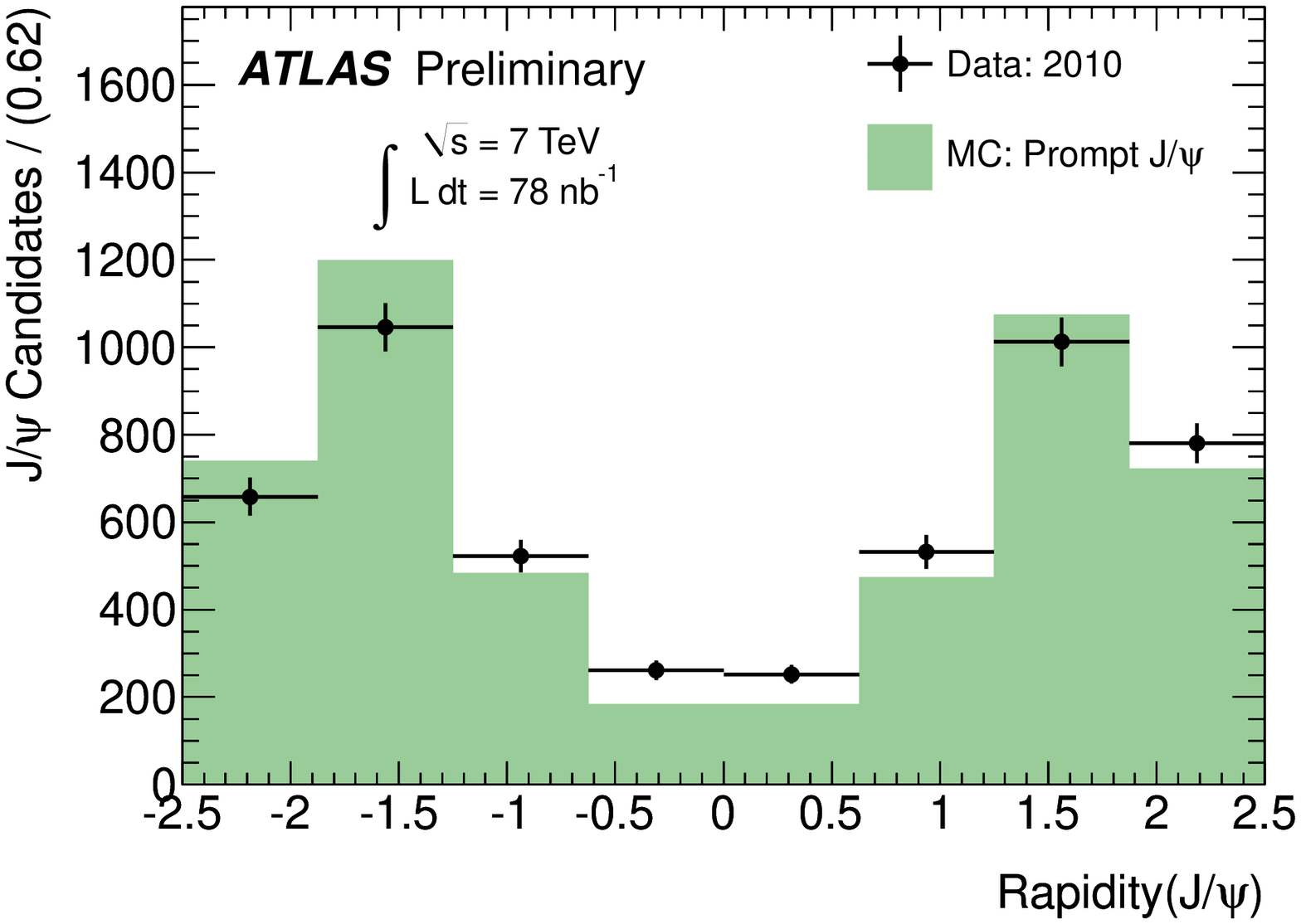}
    \caption{The transverse momentum distribution (left) and rapidity distribution of background-subtracted $J/\psi$ candidates, and a comparison to prompt $J/\psi$ Monte Carlo simulation.
      \label{fig:jpsikin}}
  \end{center}
\end{figure}

An unbinned maximum likelihood fit was used to fit the $J/\psi$ signal. The signal is modelled with a Gaussian distribution, while the background is modelled with a linear function,
with the means and resolutions calculated on an candidate-by-candidate basis and the errors calculated from the covariance matrix of the vertex fit.
The invariant mass of the selected $J/\psi$ candidates is shown in Figure~\ref{fig:jpsimass}, split by the pseudorapidities of the two muons. One can see that the level 
of background changes with pseudorapidity, largely due to the $p_T$-dependence of the combinatorial background. 
Backgrounds come dominantly from decays in flight of pions and kaons at low-$p_T$, but other significant contributions come from heavy flavour decays and to a lesser extent,
Drell-Yan processes. The mass resolution of the $J/\psi$ varies with muon pseudorapidity. This is due largely to increased traversed material and  
shorter effective trajectories in the magnetic field at higher pseudorapidities which degrade the muon momentum resolution, and thus the $J/\psi$ mass resolution from $34\pm 1$~MeV to 
$79\pm 2$~MeV within the three cases considered, but the effect is well-reproduced in Monte Carlo simulation. The $J/\psi$ fitted mass mean is in agreement with
world averages in all cases. The background-subtracted kinematic distributions of the $J/\psi$ candidates within $2.7<M(\mu^+\mu^-)<3.5$ are shown in Figure~\ref{fig:jpsikin} 
in comparison to prompt $J/\psi$ Monte Carlo simulation, with the overall normalisation taken from the signal fit in data.

\begin{figure}[htbp]
  \begin{center}
    \includegraphics[width=0.4\textwidth]{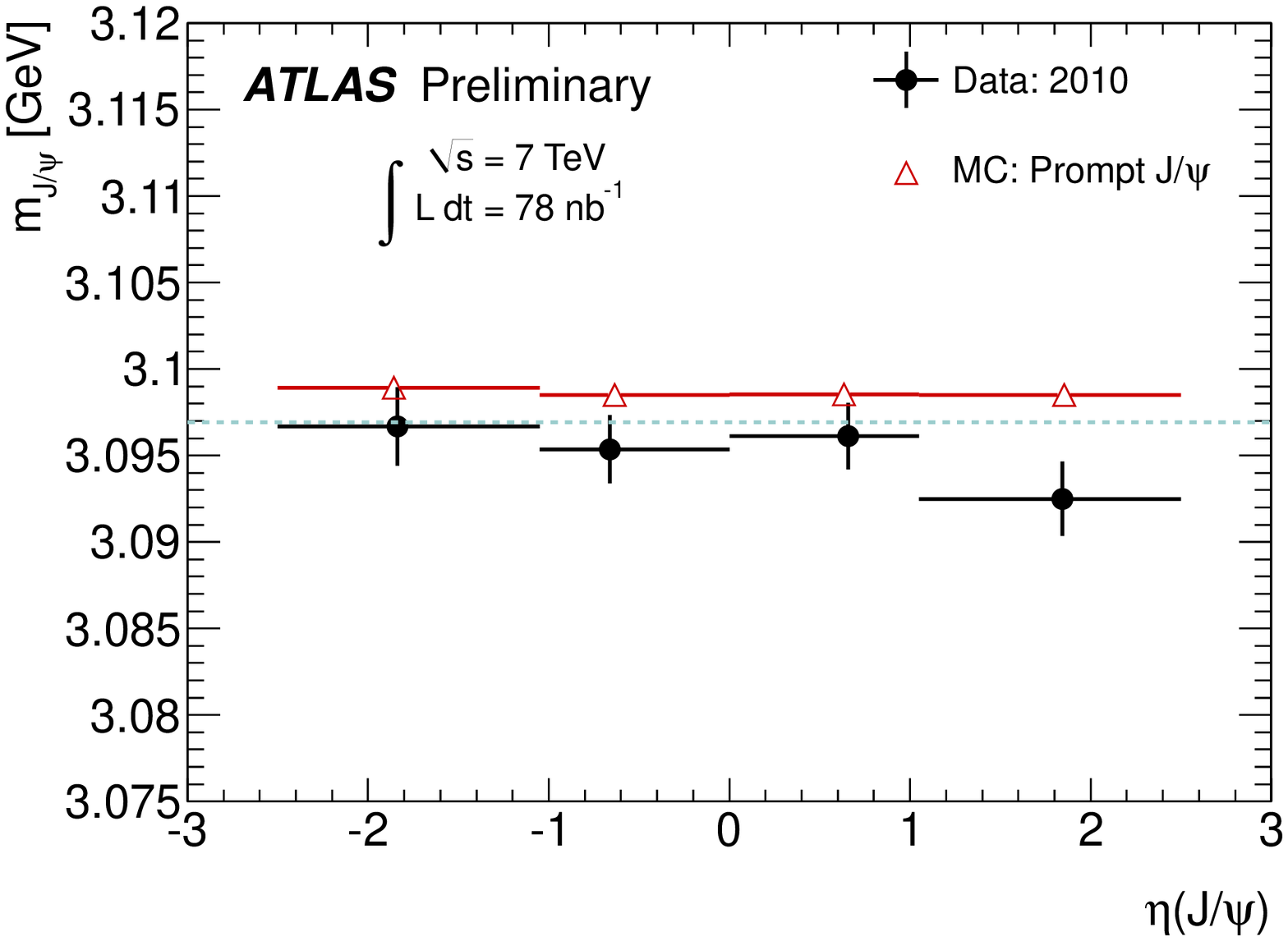}
    \includegraphics[width=0.4\textwidth]{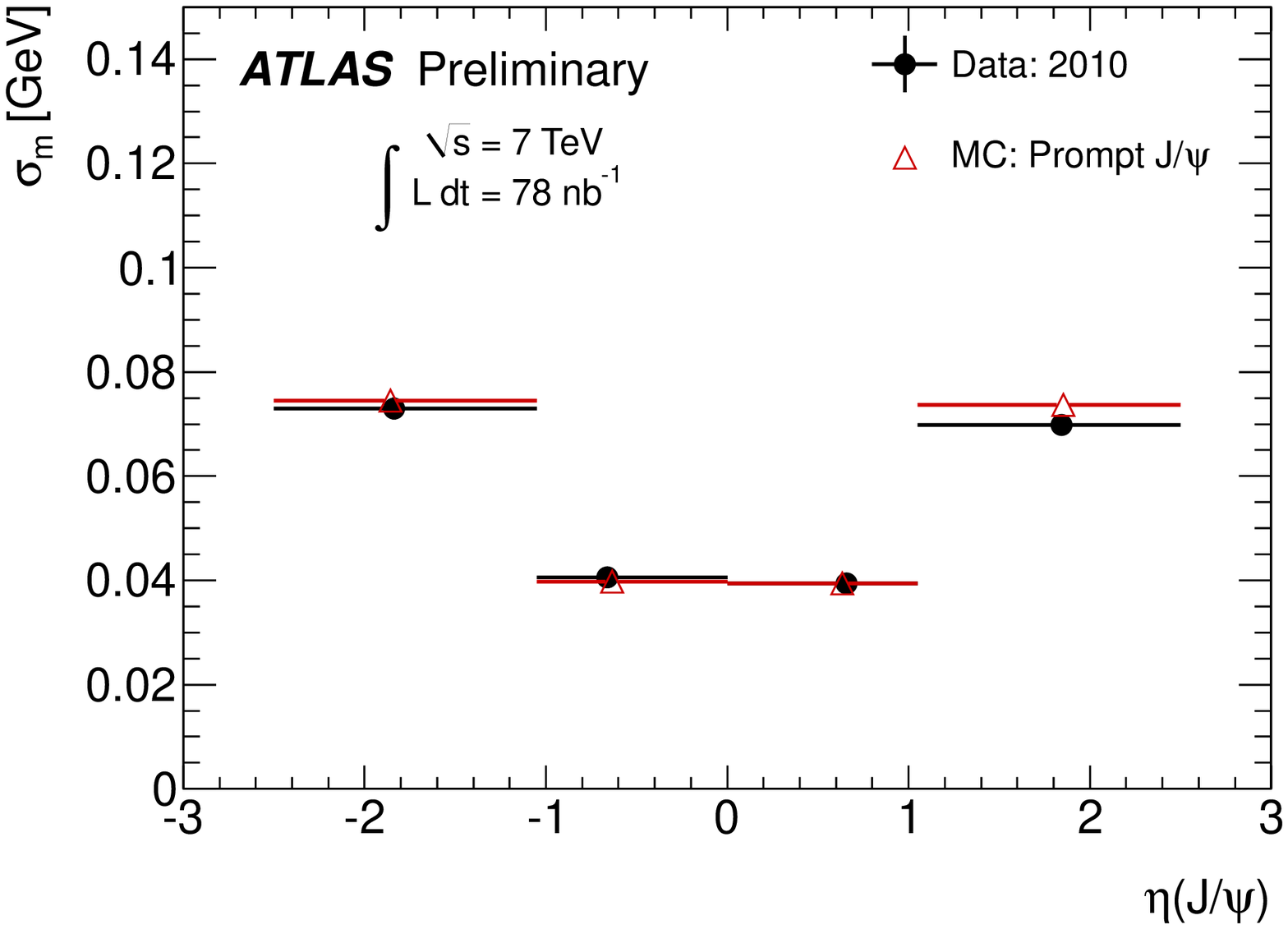}
    \caption{Reconstructed mass mean position (left) and mass resolution (right) of the $J/\psi$ as a function of the pseudo-rapidity of $J/\psi$ 
      and comparison with Monte Carlo simulation. No significant bias is seen at this time. 
      \label{fig:jpsiperf}}
  \end{center}
\end{figure}

The reconstructed mass and mass resolution are expected to be sensitive to detector misalignment and mismodelling effects. With this in mind, the results of fits to the $J/\psi$ 
resonance (and muon kinematics) were used to search for shifts (and correlations) in mean position and resolution in particular kinematic ranges. The results of such a study on
$J/\psi$ pseudorapidity are shown in Figure~\ref{fig:jpsiperf}. No significant biases have been seen to date, implying there are no large-scale mismodelling or misalignment effects
of the detector which the $J/\psi$ is sensitive to at these integrated luminosities, which bodes well for the physics programme at ATLAS.


\vspace{-0.6cm}
\section*{References}
\begin{thebibliography}{99}

\bibitem{ATLAS} ATLAS Collaboration, {\em The ATLAS Experiment at the CERN Large Hadron Collider}, JINST {\bf 3} (2008) S08003;
  L. Evans and P. Bryant, {\em LHC machine}, JINST {\bf 3} (2008) S08001.

\bibitem{DmesonCONF} ATLAS Collaboration, {\em $D^{*}$ mesons reconstruction in pp collisions at $\sqrt{s} = 7$~TeV}, 
  Tech.\ Rep., ATLAS-CONF-2010-034, CERN, Geneva, Jul 2010.

\bibitem{jpsiCONF1} ATLAS Collaboration, {\em First observation of the $J/\psi\to\mu^+\mu^-$ resonance in
  ATLAS pp collisions at $\sqrt{s} = 7$~TeV}, Tech.\ Rep., ATLAS-CONF-2010-045, CERN, Geneva, Jul 2010.
  
\bibitem{jpsiCONF2} ATLAS Collaboration, {\em $J/\psi$ performance of the ATLAS Inner Detector},
  Tech.\ Rep., ATLAS-CONF-2010-078, CERN, Geneva, Jul 2010.

\end{thebibliography}
\end{document}